\documentclass[12pt,a4paper]{article}

\usepackage{amsmath,amssymb,graphicx,epsfig}

\begin{document}

\title{\bf On the exploitability of thermo-charged capacitors}

\author{Germano D'Abramo\vspace{0.2cm}\\
{\small Istituto Nazionale di Astrofisica,}\\ 
{\small Via Fosso del Cavaliere 100,}\\
{\small 00133, Roma, Italy.}\\
{\small E--mail: {\tt Germano.Dabramo@iasf-roma.inaf.it}}}
\vspace{0.2cm}
\date{{\em Physica A} {\bf 390/6} (2011) 482 + Addendum (2013)}

\maketitle

\begin{abstract}

Recently [{\em Physics Letters A} 374 (2010) 1801] the concept of vacuum 
capacitor spontaneously charged harnessing the heat from a single 
thermal reservoir at room temperature has been introduced, along with a 
mathematical description of its functioning and a discussion on the main 
paradoxical feature that seems to violate the Second Law of 
Thermodynamics. In the present paper we investigate the theoretical and 
practical possibility of exploiting such a thermo-charged capacitor as 
voltage/current generator: we show that if very weak provisos on the 
physical characteristics of the capacitor are fulfilled, then a non-zero 
current should flow across the device, allowing the generation of 
potentially usable voltage, current and electric power out of a single 
thermal source at room temperature. Preliminary results show that the 
power output is tiny but non-zero.\\

\noindent {\bf PACS (2010):} 79.40.+z, 67.30.ef, 65.40.gd, 84.32.Tt

\end{abstract}

\section{Introduction}

In a recent paper~\cite{io}, the author introduces the concept of vacuum 
capacitor spontaneously charged harnessing the heat absorbed from a 
single thermal reservoir at room temperature. It is called for brevity 
{\em thermo-charged capacitor}. Further, he presents a mathematical 
description of its basic functioning.

In the same paper, the author shows that, when the tools of the 
Classical Thermodynamics, e.g. the Clausius entropy variation analysis, 
are applied to the process, a paradox seems to arise: the macroscopic 
behavior of a thermo-charged capacitor appears to violate the Clausius 
formulation of the Second Law of Thermodynamics~\cite{io}.

As a matter of fact, such a result should not be seen as so weird. 
Although no experimental violation has been claimed to date, over the 
past 10-15 years an unparalleled number of challenges has been proposed 
against the status of the Second Law of Thermodynamics. During this 
period, more than 50 papers have appeared in the refereed scientific 
literature (see, for example, references~\cite{bib1,bib2,bib3, 
bib4,bib5,bib6,bib7,bib8,bib9,bib10,bib11,bib12,bib13,bib14,bib15,bib16, 
bib17,bib18,bib19}), together with a monograph entirely devoted to this 
subject~\cite{cs}. Moreover, during the same period two international 
conferences on the limit of the Second Law were also 
held~\cite{sh1,sh2}.

The general class of recent challenges~\cite{cs,bib16,bib19} spans 
plasma~\cite{bib15}, chemical~\cite{bib18}, gravitational~\cite{bib8} 
and solid state physics~\cite{bib17,bib17b,bib19}. Currently, all these 
approaches appear immune to standard Second Law defenses (for a 
compendium of the classical defenses, see~\cite{eanor}) and several of 
them account laboratory corroboration of their underlying physical 
processes.

In the present paper we investigate the theoretical and practical 
possibility of exploiting thermo-charged capacitors as voltage/current 
generators. In Section~2 we review the setup of the thermo-charged 
capacitor and its mathematical model, extending that presented 
in~\cite{io}. Here we drop an apparently crucial simplification made 
in~\cite{io} and show that the physical process presented in~\cite{io} 
is robust against such simplification.

In Section~3 we show that if very weak provisos on the physical 
characteristics of the capacitor are fulfilled, then a non-zero current 
should flow across the device, allowing the generation of potentially 
usable voltage, current and electric power out of a single thermal 
source at room temperature. Preliminary results show that the power 
output is tiny but non-zero.

If it were possible to experimentally and unambiguously achieve such 
results, then we would have an experimental violation of the Second Law 
in the Kelvin-Planck formulation, which parallels the alleged violation 
in its Clausius form discussed in~\cite{io}. In other words, we could 
have a sort of {\it thermo-voltaic cell}.

\section{Thermo-charged spherical capacitor}

\begin{figure}[t]

\begin{center}
\begin{picture}(150,150)
\setlength{\unitlength}{0.5cm} \thicklines \put(4,6){\circle{1}}
\put(4,6){\circle{1.08}} \put(4,6){\circle{1.1}}
\put(4,6){\oval(4,4)}
\put(4,5.5){\line(0,-1){4}}\put(4.05,1.5){\vector(-1,0){0.6}}
\put(2.06,1.5){\vector(1,0){0.6}} \put(2.06,5){\line(0,-1){3.5}}
\put(2.9,0.8){{\scriptsize $V$}} \put(1.6,1){{\scriptsize
$\ominus$}} \put(4.1,1){{\scriptsize $\oplus$}}
\put(6.5,8.5){{\scriptsize Opaque case (with $\phi_{ext}$)}}
\put(7.1,7.8){{\scriptsize Externally insulated}}
\put(2.06,1.5){\vector(1,0){0.6}} \put(0,8){{\scriptsize Vacuum}}
{\thinlines\put(5.5,7.65){\vector(1,1){0.8}}
\put(1.5,7.65){\vector(1,-1){1.2}}} \put(4,6.1){{\scriptsize $a$}}
\put(5,5.1){{\scriptsize $b$}}
{\thinlines\put(4,6){\line(1,0){0.5}}}
{\thinlines\put(4,6){\line(1,-1){1.6}}} \put(7,5){{\scriptsize Room
temperature $T\simeq 298\,$K}} {\thinlines
\put(3.5,6){\vector(-1,0){3}}} \put(-3.5,6){{\scriptsize Ag--O--Cs
layer}} {\thinlines \put(3,7.3){\vector(1,-1){0.8}}} {\thinlines
\put(3.8,6.5){\vector(1,1){0.8}}} \put(2.7,7.4){\scriptsize $h\nu$}
\put(4.5,7.4){\scriptsize $e^-$}


\end{picture}
\end{center}
\caption{Scheme of the thermo-charged spherical capacitor.}
\label{fig1}
\end{figure}
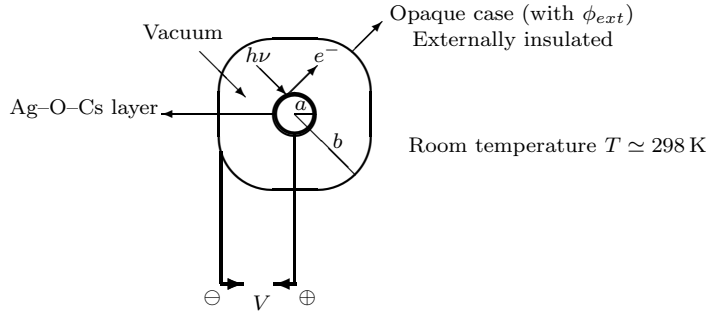

In Fig.~\ref{fig1} a sketched section of the vacuum spherical capacitor 
is shown (see also~\cite{io}). The outer sphere has radius $b$ and it is 
made of metallic material with relatively high work function 
($\phi_{ext} > 1\,$eV). The inner sphere has radius $a$ and it is made 
of the same conductive material as the outer one, but it is coated with 
a layer of semiconductor Ag--O--Cs, which has a relatively low work 
function ($\phi_{in}\lesssim 0.7\,$eV). It should be clear that in such 
a case the two thermionic fluxes, from each plate toward the other one, 
are different, the latter being greater than the former, at least at the 
beginning of the charging process. The capacitor should be shielded by a 
case and put at room temperature. The case must be opaque to every 
environmental electromagnetic disturbance (natural and man-made 
e.m.~waves, cosmic rays and so on) in order to avoid spurious {\em 
photo}-electric emission. Moreover, the outer plate should be externally 
insulated, in order to prevent its outward thermionic emission and the 
inter-plate space must be under extreme vacuum (UHV).

All the electrons emitted by the inner sphere, due to thermionic 
emission at room temperature, are collected by the outer (very low 
emitting) sphere, creating a macroscopic difference of potential $V$. At 
first, such a process is unbalanced, being the flux from the inner 
sphere greater than that from the outer sphere, but later, with the 
increase of potential $V$, the inward and the outward flux should tend 
to balance each other.

A crucial point for what follows is the behavior of the contact surface 
between the Ag--O--Cs layer and the conductive material of the inner 
sphere.

\begin{figure}[t]
\begin{center}
\includegraphics[height=5cm,width=6cm]{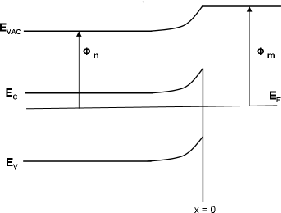}
\end{center}
\caption{Band profiles of semiconductor (n) -- metal (m) junction at 
equilibrium; $\phi_n$ and $\phi_m$ (with $\phi_n < \phi_m$) are the work 
functions of semiconductor and metal, respectively. ($x=0$ indicates the 
contact between surfaces.)}
\label{fig1b}
\end{figure}

The contact surface between the inner metallic plate and the Ag--O--Cs 
layer is a well-known Schottky junction (metal/n--type semiconductor). 
When two materials (in our case, a metal and a semiconductor) are 
physically joined, so as to establish a uniform chemical potential, that 
is a single Fermi level, some electrons are transferred from the 
material with the lesser work function $\phi_1$ (Ag--O--Cs) to the 
material with the greater work function $\phi_2$ (metal). As a result, a 
contact potential $V_c$ is established such that $eV_c=\phi_2-\phi_1$. 
The junction is thus the region where, at equilibrium, a balance between 
bulk electrostatic and diffusive (thermally driven) forces is attained.
The energy band profiles of semiconductor-metal junction at equilibrium 
are shown in Fig.~\ref{fig1b}. The feature which counts for the 
functioning of our device is the fact that the energy level of the 
vacuum for Ag--O--Cs (and that for the metallic plate) is preserved far 
from the depletion region~\cite{ueb}.

This means that whenever an electron is extracted from the Ag--O--Cs 
layer to the vacuum (toward the outer sphere), and this is made always 
at the cost of $\approx 0.7\,$eV for what is said above, Ag--O--Cs layer 
starts to charge up positively; hence, a sort of `external' reverse bias 
starts to form across the junction and a tiny current of electrons 
begins to flow from the metallic inner sphere to the Ag--O--Cs layer in 
order to re-establish the equilibrium (constant contact potential).

Such a current flow is known as {\em reverse bias leakage current} 
(RLC). Its physical generating mechanisms are variegated and complex, 
and the amplitude is influenced by many factors like thickness of the 
depletion region, temperature, cross-sectional area and impurities of 
the junction, and so on. Also the amplitude of the reverse bias 
influences the intensity of RLC, although when the reverse bias is below 
the {\em breakdown voltage} of the junction (usually tens or hundreds of 
volts), the current changes slowly with bias.
 
It depends on the used material, but usually the {\em reverse leakage 
current density} $j_0$, namely RLC per unit surface, spans\footnote{See, 
for example, \cite{oyama,rossi,hsu,dann} for some specific types of 
Schottky and n--p junctions. Values greater than $10^{-6}$A/cm$^2$ and 
lower than $10^{-9}$A/cm$^2$ are also possible with the same reverse 
bias, depending on the materials, preparation, junction impurities and 
surface treatments. Usually, applied researchers and industry desire to 
lower the reverse leakage current in order to exalt the rectifying 
properties of the junction for electrical and electronic applications. 
Here, instead, we have opposite needs.} from $10^{-6}$A/cm$^2$ to 
$10^{-9}$A/cm$^2$ for reverse biases of the order of volts or tenths of 
volts. As will be clear later, if we suitably increase the contact 
surface between the inner metallic sphere and the Ag--O--Cs layer, 
namely increase the surface of the inner sphere $S_a$, then the RLC can 
be made of the order of $10^{-3}$A or greater, RLC being equal to 
$j_0S_a$. We can also be favored by junction impurities: they usually 
transform rectifying junctions into ohmic ones. All these issues will be 
treated more thoroughly in Section~3.

\begin{figure}[t]
\begin{center}
  \includegraphics[height=7cm]{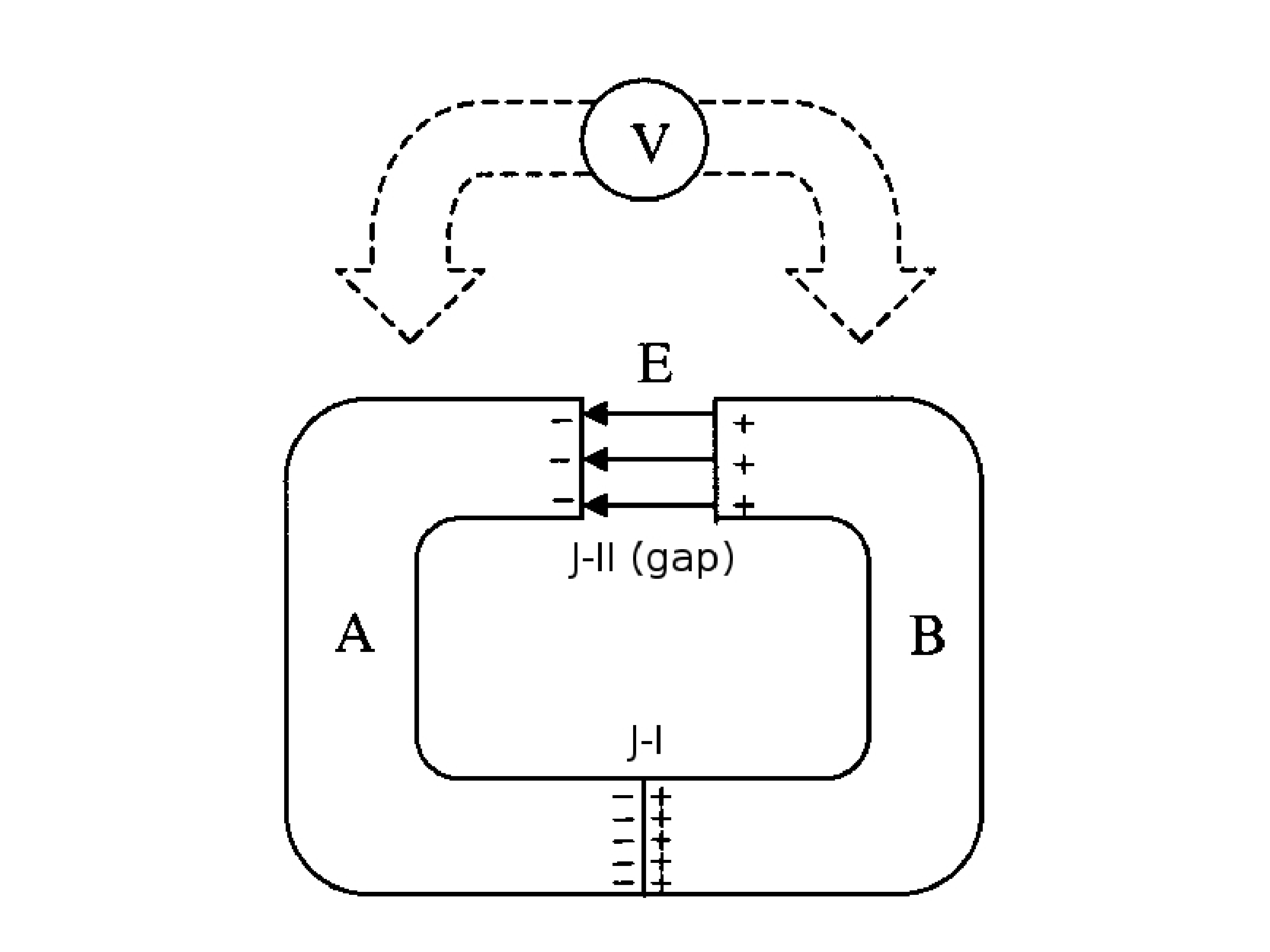}
\end{center}
\caption{Circuit of two connected materials A and B in vacuum. This scheme 
holds for metal to metal and metal to semiconductor junctions. Work 
functions are such that $\phi_A > \phi_B$. J-I is the physical junction 
while J-II is the gap (free ends). (Adapted from~\cite{uwp})}
\label{fig10}
\end{figure}

In almost all textbooks it is said that a voltage drop $V$ builds up 
not only across the contact surface (contact potential), but 
instantaneously also between the surfaces at the free ends of the materials 
as soon as they are joined at one end (see Fig.~\ref{fig10}), where charges 
also accumulate (to the author's knowledge, no textbook or published paper 
on the subject clearly explains where these charges come from). This voltage 
drop is of the same magnitude of the contact potential $V_c$. All this is 
usually explained by appealing to a supposed straightforward application of 
the Kirchhoff's second (loop) rule.

If this were true, then it should prevent any imbalance in electron emission
between the inner and the outer sphere as soon as the two spheres are 
electrically shorted through an external resistor so as to form a
circuit similar to that in Fig.~\ref{fig10}. In that case, in order to reach 
the outer sphere, any electron escaping the inner layer would need the 
same energy needed by an electron escaping the outer sphere to reach the 
inner one. The inner electron needs an energy equal to $(\phi_{in} + 
eV)$, since it must be ejected (required energy $\phi_{in}$) and 
then it has to overcome the potential drop $V$ (energy equal to 
$eV$). The outer electron needs $\phi_{ext}$ (energy to be 
ejected). But, since $eV_c$=$eV=\phi_{ext}-\phi_{in}$, the inner 
electron needs $(\phi_{in} + eV)= \phi_{ext}$.
Such a situation would undoubtedly be one of equilibrium. 
But we have explicitly shown in~\cite{io3} that no electric field, and 
thus no voltage drop, builds up between the surfaces at the free ends of 
two materials with different work functions as soon as the materials are 
physically joined at one end. 

In~\cite{io3} we presented three arguments: the first two were more 
heuristic, the third one was more theoretical. Let us sketch here the last 
one, namely an explicit application of the {\em path-independence} law and/or 
Kirchhoff's loop rule. The physical principle at the basis of these two laws 
is the more fundamental law of conservation of energy. Conservation of energy 
demands that a test electronic charge $e$ conveyed around a closed path 
$\gamma$ in the device bulk of Fig.~\ref{fig10}, through J-I (physical 
junction) and J-II (gap) at equilibrium, must undergo zero net work from 
{\em all} the forces present along the path. Mathematically, we must have,

\begin{equation}
\oint_{\gamma} d\textrm{W}_{ext}=0.
\label{eq1ad}
\end{equation}

At equilibrium, the only two regions where forces are allowed to be 
non-zero are the J-I and J-II regions. An electric field elsewhere in A 
and B (other than in the contact region) would generate a current, which 
contradicts the assumption of equilibrium. When the test charge $e$ 
crosses J-I, it is subject to the built-in electric field force 
$e\textrm{\bf E}_{bi}$ and to the diffusion force $\textrm{\bf 
F}_{diff}$. This ``force'' is a thermally driven force and it is 
responsible for the establishment of the contact potential at J-I.  We 
know that at equilibrium $e\textrm{\bf E}_{bi}=-\textrm{\bf F}_{diff}$ 
and that $\textrm{\bf F}_{diff}$ is different from zero and constantly 
present, otherwise $\textrm{\bf E}_{bi}$ would soon drop to zero, thus,

\begin{equation}
0=\oint_{\gamma} d\textrm{W}_{ext}=\int_{\textrm{\bf J-I}}
(e\textrm{\bf E}_{bi}+\textrm{\bf F}_{diff})\cdot d{\vec \gamma} + 
\int_{\textrm{\bf J-II}}d\textrm{W}_{ext}=0+
\int_{\textrm{\bf J-II}}d\textrm{W}_{ext}.
\label{eq2ad}
\end{equation}

In the J-II gap there are no diffusion forces, since it is a vacuum gap, 
and eventually we have,

\begin{equation}
0=\int_{\textrm{\bf J-II}}d\textrm{W}_{ext}=\int_{\textrm{\bf J-II}}
e{\textrm{\bf E}_{\textrm{\bf J-II}}}\cdot d{\vec \gamma}=
e|\textrm{\bf E}_{\textrm{\bf J-II}}|x_{g}
\quad\to\quad |\textrm{\bf E}_{\textrm{\bf J-II}}|=0,
\label{eq3ad}
\end{equation}
where $x_{g}$ is the gap width.

We now refine the estimate of the obtainable voltage and the estimate of 
the time needed to reach such a value, taking into account not only the 
physical characteristics of the capacitor and the quantum efficiency 
curve $\eta_{in}(\nu)$ of thermionic material Ag--O--Cs (as made 
in~\cite{io}), but also the thermionic emission and the quantum 
efficiency curve $\eta_{ext}(\nu)$ of the outer sphere.

The capacitor is placed in a heat bath at room temperature and it is 
subject to the black-body radiation. Both spheres, at thermal 
equilibrium, emit and absorb an equal amount of radiation (Kirchhoff's 
law of thermal radiation), thus the amount of radiation absorbed by the 
spheres is the same as that emitted by those spheres according to the 
black-body radiation formula (Planck equation). Hence, given the room 
temperature $T$, the Planck equation provides us with the number 
distribution of photons absorbed as a function of their frequency.

According to the law of thermionic emission, the kinetic energy $K_e$ of 
an electron emitted by the material is given by the following equation,

\begin{equation}
K_e=h\nu-\phi,
\label{eq1}
\end{equation}
where $h\nu$ is the energy of the photon with frequency $\nu$ ($h$ is 
the Planck constant) and $\phi$ is the work function of the material. 
Thus, only the tail of the Planck distribution of the absorbed photons, 
with frequency $\nu>\nu_0=\phi/h$, can contribute to thermionic emission.

An electron emitted by the inner sphere, thanks to the absorption of a 
photon of frequency $\nu_1$, is able to reach the outer sphere with zero 
velocity only when,

\begin{equation}
eV =h\nu_1-\phi_{in},
\label{eq2}
\end{equation}
where $V$ is the inter-sphere voltage reached so far. On the other hand, 
an electron of the outer sphere does not have to overcome an opposed voltage 
in order to reach the inner sphere, and thus the following holds,

\begin{equation}
h\nu_2=\phi_{ext}.
\label{eq2b}
\end{equation}

Hence, we have the following useful relations, 

\begin{equation}
\nu_1=\frac{eV +\phi_{in}}{h},
\label{eq3}
\end{equation}

\begin{equation}
\nu_2=\frac{\phi_{ext}}{h},
\label{eq3b}
\end{equation}
which give the minimum frequencies of radiation with enough energy to move
an electron from one sphere to the other.

The total number of photons per unit time $F_{p}$ with energy greater 
than or equal to $h\nu_1$, emitted and absorbed in thermal equilibrium 
by the inner sphere is given by the Planck equation,

\begin{equation}
F_{p}=\frac{2\pi S_a}{c^2}\int_{\nu_1}^\infty
\frac{\nu^2 d\nu}{e^{\frac{h\nu}{kT}}-1},
\label{eq4}
\end{equation}
where $S_a$ is the inner sphere surface area, $c$ is the speed of light, 
$k$ is the Boltzmann constant and $T$ the room temperature.

If $\eta_{in}(\nu)$ is the quantum efficiency (or quantum yield) curve 
of the Ag--O--Cs thermionic layer, then the number of electrons per unit 
time $F_{in}$ emitted by the inner sphere towards the outer sphere with 
kinetic energy greater than or equal to $h\nu_1-\phi$, is given by,

\begin{equation}
F_{in}=\frac{2\pi\cdot 4\pi a^2}{c^2}\int_{\nu_1}^\infty
\frac{\eta_{in}(\nu)\nu^2 d\nu}{e^{\frac{h\nu}{kT}}-1},
\label{eq5}
\end{equation}
where $4\pi a^2$ is the surface area of the inner sphere.

Following the same reasoning, the number of electrons per unit time 
$F_{ext}$ emitted by the outer sphere and collected by the inner one is 
given by the analogous relation,

\begin{equation}
F_{ext}=\frac{2\pi\cdot S_{eff}}{c^2}\int_{\nu_2}^\infty
\frac{\eta_{ext}(\nu)\nu^2 d\nu}{e^{\frac{h\nu}{kT}}-1}.
\label{eq5b}
\end{equation}
 
In this last relation it is not easy to define the multiplicative factor 
related to the surface area: before the charging process starts, $S_{eff}$ is 
equal to the inner sphere surface area $4\pi a^2$, but as soon as the 
inner sphere charges up positively, its effective surface area increases 
due to the electrostatic focusing effect (similar to the gravitational 
focusing effect). Moreover, it is not easy to mathematically model such 
a phenomenon since the effective area of the inner sphere depends on the 
velocity of the single electron flying toward it. Here we decide to be 
extremely conservative and choose $S_{eff}$ equal to the surface area of 
the outer sphere. As a matter of fact, the following results are 
practically independent of the choice of any reasonable value of 
$S_{eff}$.

For a vacuum spherical capacitor, the voltage between the spheres $V$ 
and the charge on each sphere $Q$ are linked by the following well-known 
equation,

\begin{equation}
V=\frac{Q}{4\pi \epsilon_0}\frac{b-a}{ab}.
\label{eq6}
\end{equation}

Now, we derive the differential equation which governs the process of 
thermo-charging. In the interval of time $dt$ the charge collected by 
the outer sphere is given by,

\begin{equation}
dQ=e(F_{in}-F_{ext}) dt=\frac{2\pi e}{c^2}\Biggl(4\pi a^2\int_{\frac{eV(t) +
\phi_{in}}{h}}^\infty \frac{\eta_{in}(\nu)\nu^2
d\nu}{e^{\frac{h\nu}{kT}}-1} - 4\pi b^2\int_{\frac{\phi_{ext}}{h}}^\infty
\frac{\eta_{ext}(\nu)\nu^2 d\nu}{e^{\frac{h\nu}{kT}}-1}\Biggr)dt, 
\label{eq7}
\end{equation}
where we make use of eqs.~(\ref{eq3}) and~(\ref{eq3b}) for $\nu_1$ and $\nu_2$, 
and $V(t)$ is the voltage at time $t$. Thus, through the differential form of 
eq.~(\ref{eq6}), we have,

\begin{equation}
dV(t) = \frac{2\pi e}{\epsilon_0
c^2}\Biggl(\frac{a(b-a)}{b} \int_{\frac{eV(t) + \phi_{in}}{h}}^\infty
\frac{\eta_{in}(\nu)\nu^2 d\nu}{e^{\frac{h\nu}{kT}}-1} - \frac{b(b-a)}{a}\int_{\frac{\phi_{ext}}{h}}^\infty
\frac{\eta_{ext}(\nu)\nu^2 d\nu}{e^{\frac{h\nu}{kT}}-1}\Biggr)dt,
\label{eq8}
\end{equation}
or

\begin{equation}
\frac{dV(t)}{dt}= \frac{2\pi e}{\epsilon_0
c^2}\Biggl(\frac{a(b-a)}{b}\int_{\frac{eV(t) + \phi_{in}}{h}}^\infty
\frac{\eta_{in}(\nu)\nu^2 d\nu}{e^{\frac{h\nu}{kT}}-1} -
\frac{b(b-a)}{a}\int_{\frac{\phi_{ext}}{h}}^\infty
\frac{\eta_{ext}(\nu)\nu^2 d\nu}{e^{\frac{h\nu}{kT}}-1}\Biggr).
\label{eq9}
\end{equation}

Since our aim is to maximize the value of $V$, we have to choose $a$ and 
$b$ such that they maximize the geometrical factor $a(b-a)/b$. The 
rightmost integral of eq.~(\ref{eq9}) has a smaller value with respect 
to the leftmost one by some orders of magnitude, at least at the 
beginning of the charging process, hence what counts for the 
maximization of $V$ is the maximization of the factor $a(b-a)/b$ alone.

It is not difficult to see that the maximum is reached when $a=b/2$. So 
we have,

\begin{equation}
\frac{dV(t)}{dt}= \frac{\pi e b}{2\epsilon_0
c^2} \Biggl(\int_{\frac{eV(t) + \phi_{in}}{h}}^\infty \frac{\eta_{in}(\nu)\nu^2
d\nu}{e^{\frac{h\nu}{kT}}-1} - 4\int_{\frac{\phi_{ext}}{h}}^\infty
\frac{\eta_{ext}(\nu)\nu^2 d\nu}{e^{\frac{h\nu}{kT}}-1}\Biggr).
\label{eq10}
\end{equation}

In the rest of this Section we provide a numerical solution of the above 
differential equation for the practical case of inner sphere coated with 
a layer of Ag--O--Cs ~\cite{somm1, somm2, bates}. To do that we need to 
adopt an approximation, however: the approximation consists in the 
adoption of a constant value for the functions $\eta (\nu)$, a sort of 
suitable mean values $\overline{\eta}$.

The differential equation~(\ref{eq10}) thus becomes,

\begin{equation}
\frac{dV(t)}{dt}= \frac{\pi e b}{2\epsilon_0
c^2}\Biggl(\overline{\eta}_{in}\int_{\frac{eV(t) + 
\phi_{in}}{h}}^\infty \frac{\nu^2 d\nu}{e^{\frac{h\nu}{kT}}-1} -
4\overline{\eta}_{ext}\int_{\frac{\phi_{ext}}{h}}^\infty
\frac{\nu^2 d\nu}{e^{\frac{h\nu}{kT}}-1}\Biggr).
\label{eq11}
\end{equation}

A straightforward variable substitution in the integrals of 
eq.~(\ref{eq11}) allows to write the equation in its final simplified form,

\begin{equation}
\frac{dV(t)}{dt}= \frac{\pi e b}{2\epsilon_0
c^2}\biggl(\frac{kT}{h}\biggr)^3 \Biggl(\overline{\eta}_{in}\int_{\frac{eV(t) +
\phi_{in}}{kT}}^\infty \frac{x^2 dx}{e^{x}-1} - 
4\overline{\eta}_{ext}\int_{\frac{\phi_{ext}}{kT}}^\infty
\frac{x^2 dx}{e^x-1}\Biggr).
\label{eq12}
\end{equation}

\begin{figure}[t]
\centerline{\includegraphics[width=8cm]{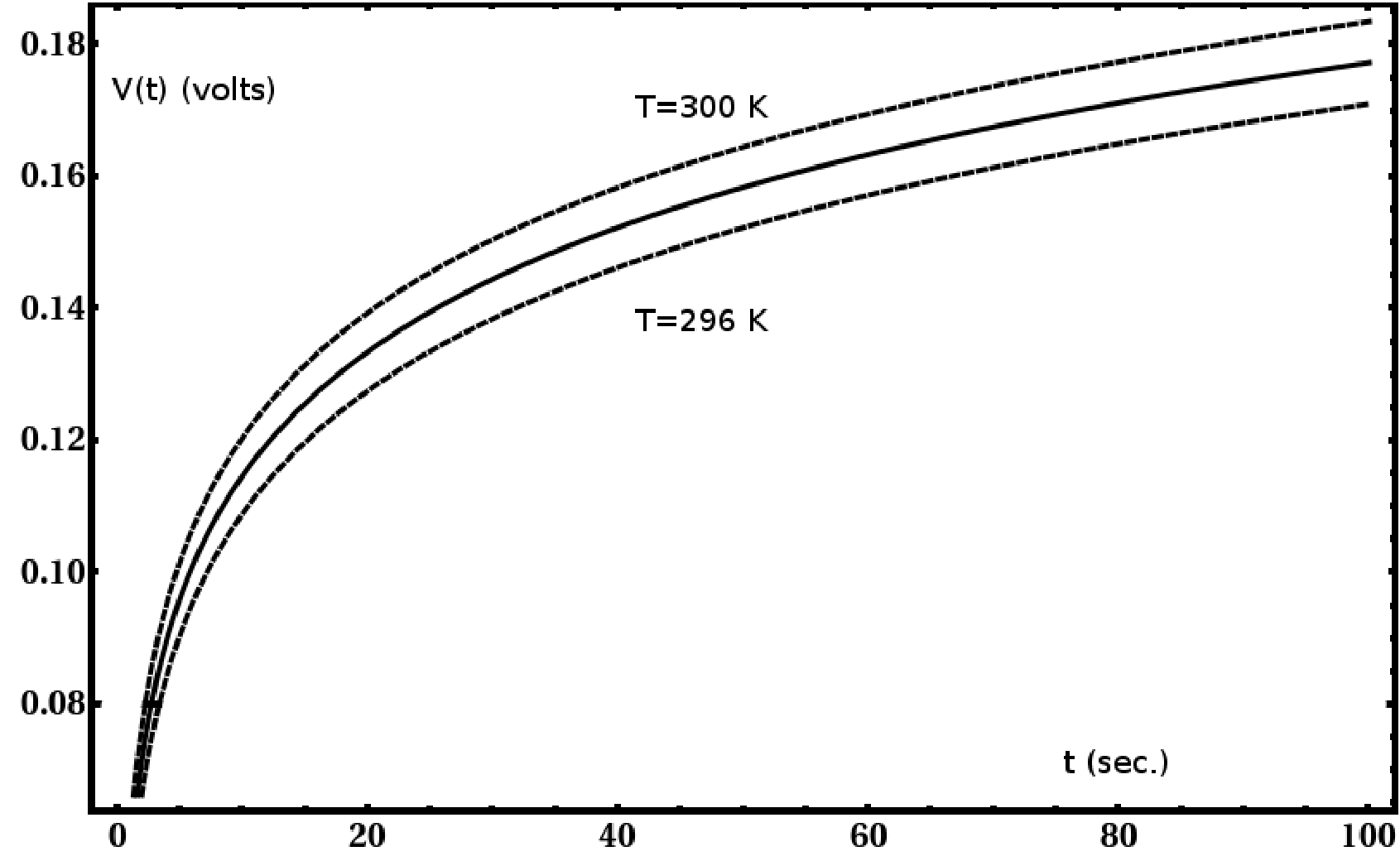}
\includegraphics[width=8cm]{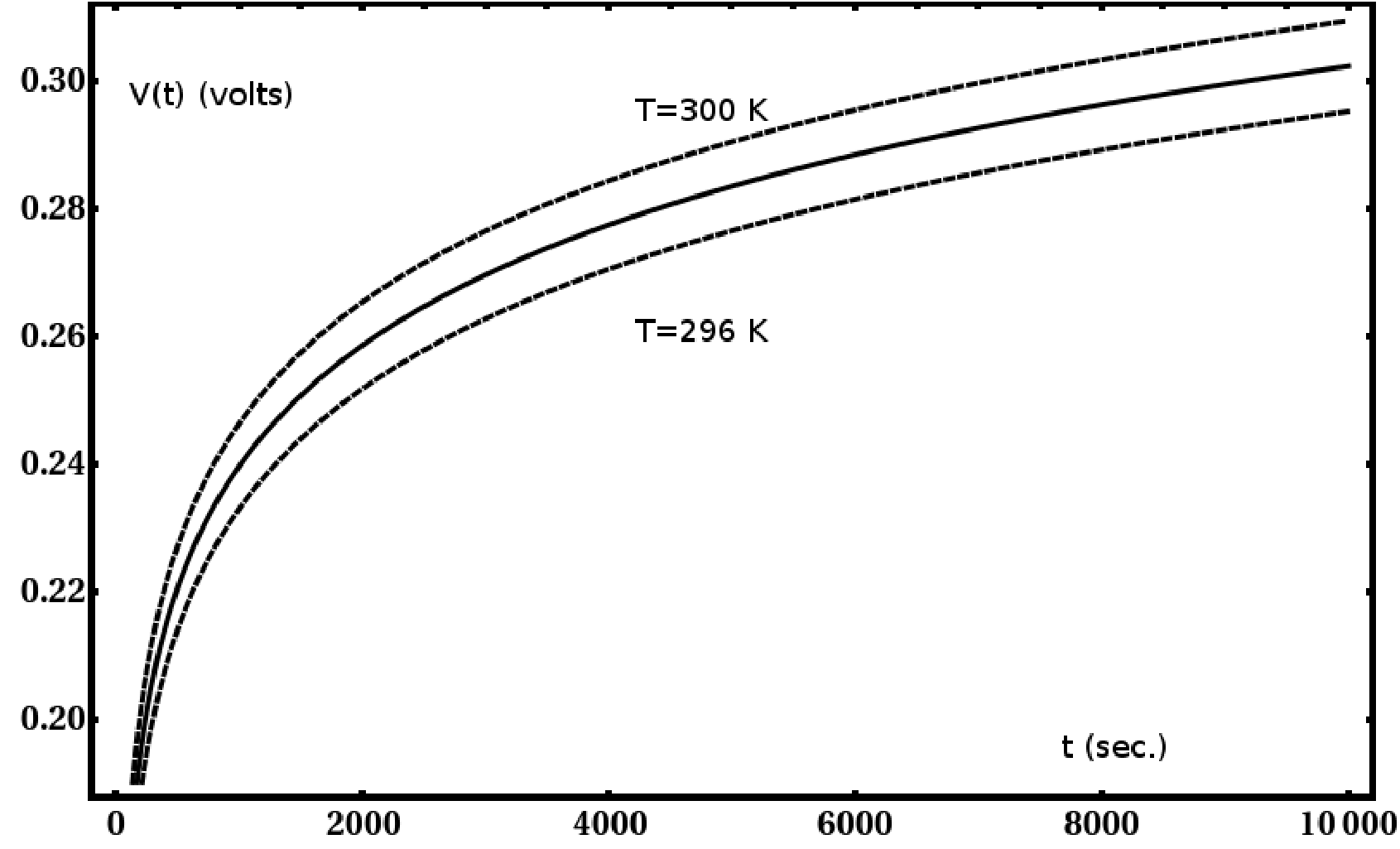}}
\caption{Thermo-charging profiles for the spherical capacitor described in 
the text ($\phi_{in}=0.7\,$eV, $\phi_{ext}=4.0\,$eV, $b=0.2\,$m, $T=298\,$K, 
$\overline{\eta}_{in}=10^{-5}$ and $\overline{\eta}_{ext}=1$). These two plots 
show with different ranges in time-scale the behavior of $V(t)$. Charging 
profiles for $T=300\,$K and for $T=296\,$K are also shown.}
\label{fig2}
\end{figure}

\begin{figure}[t]
\centerline{\includegraphics[width=8cm]{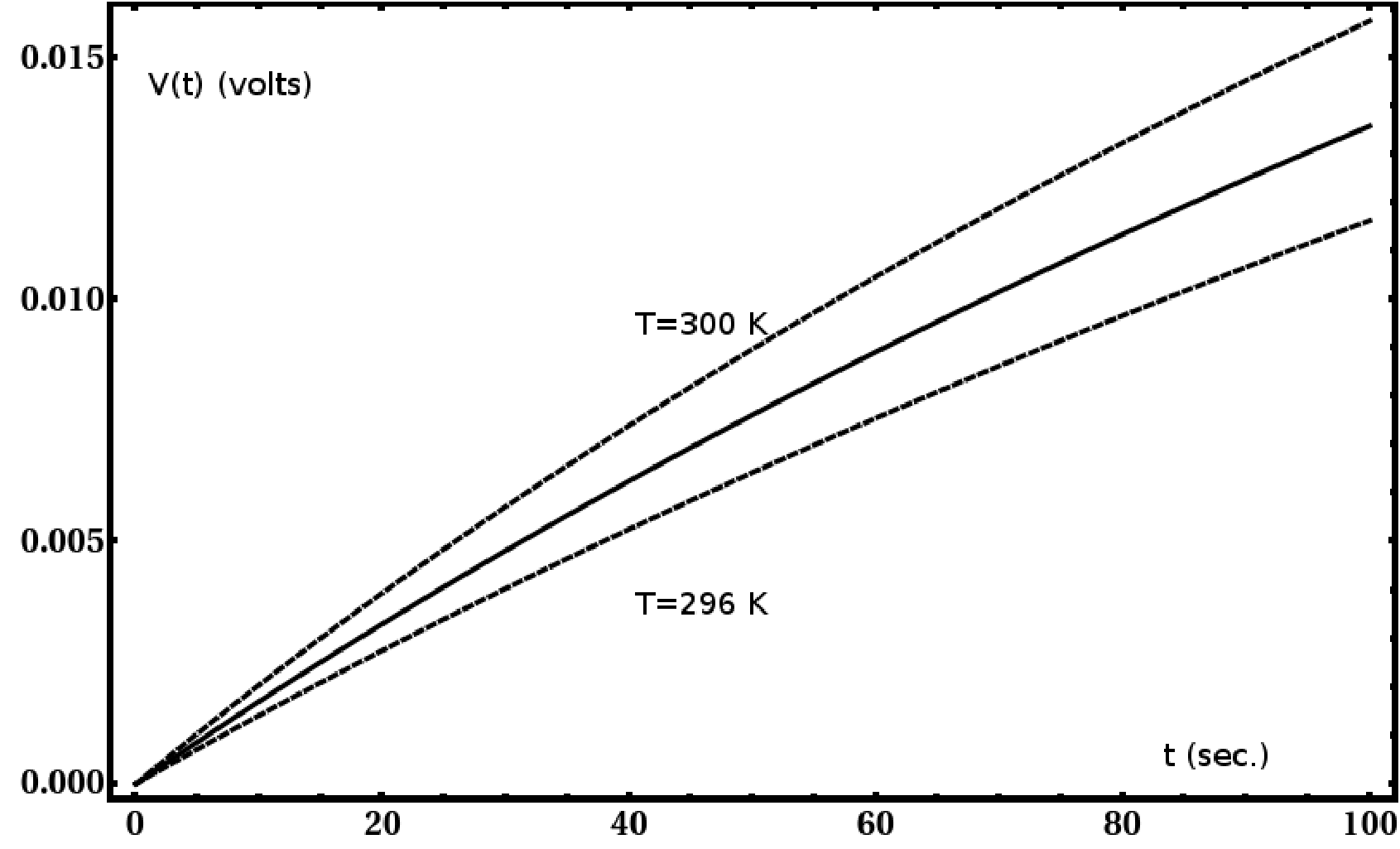}
  \includegraphics[width=8cm]{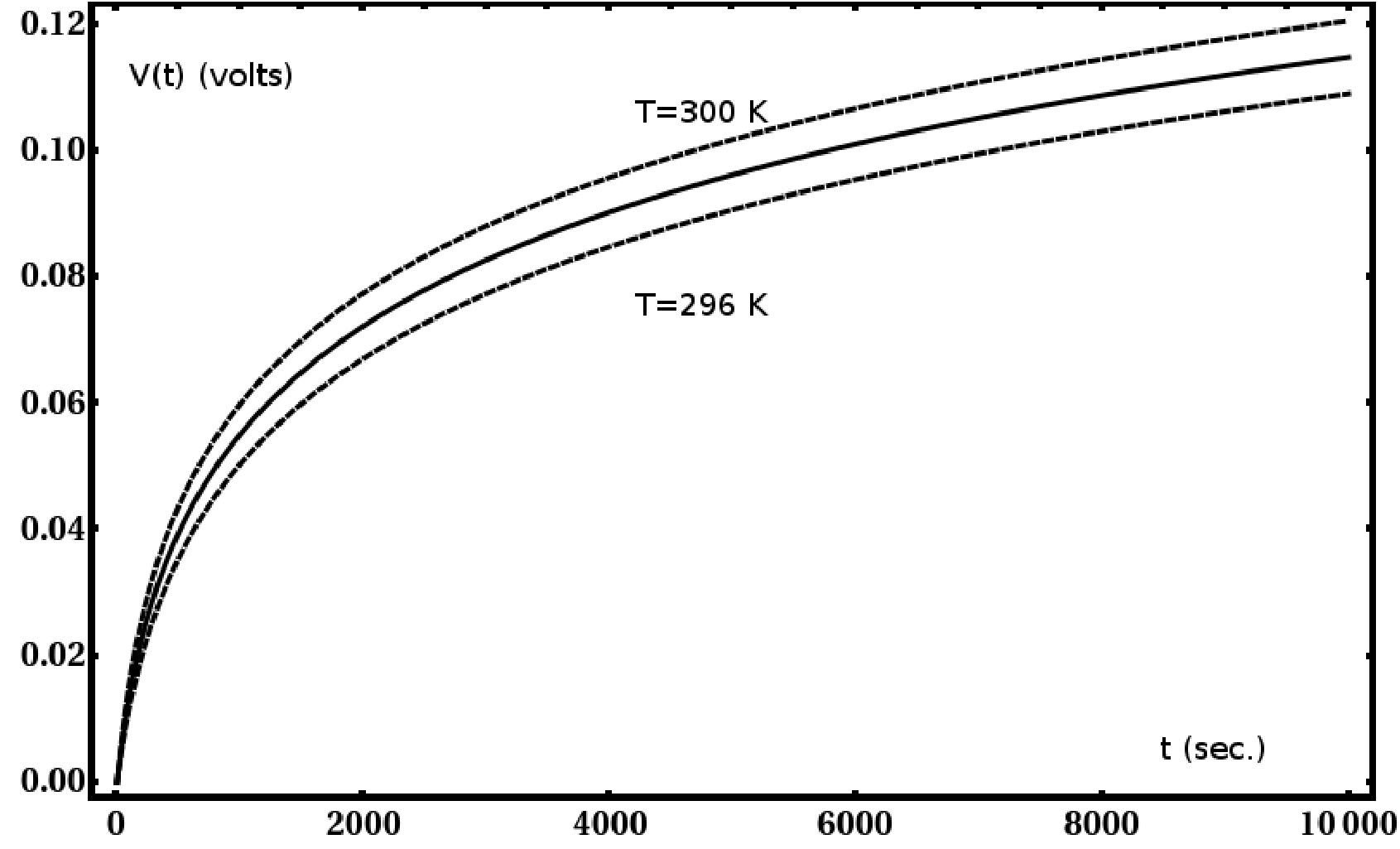}}
\caption{Thermo-charging profiles for the spherical capacitor described in 
the text with $\phi_{in}=0.7\,$eV, $\phi_{ext}=4.0\,$eV, $b=0.2\,$m, 
$T=298\,$K, $\overline{\eta}_{in}=10^{-8}$ and $\overline{\eta}_{ext}=1$. 
These two plots show with different ranges in time-scale the behavior 
of $V(t)$. Charging profiles for $T=300\,$K and for $T=296\,$K are also 
shown.}
\label{fig3}
\end{figure}

Here we provide an exemplificative numerical solution of 
eq.~(\ref{eq12}), adopting the following nominal values for $\phi_{in}$, 
$\phi_{ext}$, $b$, $T$ and $\overline{\eta}_{in}$ and 
$\overline{\eta}_{ext}$: $\phi_{in}=0.7\,$eV, $\phi_{ext}=4.0\,$eV, 
$b=0.2\,$m, $T=298\,$K, $\overline{\eta}_{in}=10^{-5}$, and 
$\overline{\eta}_{ext}=1$.

In order to make a conservative choice for the value of 
$\overline{\eta}_{in}$ we note that only black-body radiation with 
frequency greater than $\nu_0=\phi_{in}/h$ can contribute to thermionic 
emission. This means that, for the Ag--O--Cs photo-cathode, only 
radiation with wavelength smaller than $\lambda_0=hc/\phi_{in}\simeq 
1700\,$nm contributes to the emission. According to Fig.~1 
in~\cite[Bates]{bates}, the quantum yield of Ag--O--Cs for wavelengths 
smaller than $\lambda_0$ (and thus, for frequency greater than $\nu_0$) 
is always greater than $10^{-5}$.

For what concerns $\overline{\eta}_{ext}$, its value is related (as it 
is for the value of $\phi_{ext}$) to the discharging process due to 
counter emission. As a matter of fact, thermionic counter emission can 
be kept extremely low with value of $\phi_{ext}$ high compared to 
$\phi_{in}$, and also choosing suitable metallic material for the outer 
sphere, with very low $\overline{\eta}_{ext}$, namely 
$\overline{\eta}_{ext}\ll \overline{\eta}_{in}$.

There are also other possible sources of counter emission, e.~g. 
secondary electron emission~\cite{hard}, but as far as we known they are 
always smaller than the charging emission (and they can be made very 
weak with particular design, choice of material and surface texture of 
the outer sphere). In any case, counter emissions should only retard the 
achievement of the equilibrium voltage $V$, not preclude it. Any such 
delay in time can be easily modeled with the same eq.~(\ref{eq12}), 
simply changing the numerical value of $\phi_{in}$, $\phi_{ext}$, 
$\overline{\eta}_{ext}$ and $\overline{\eta}_{in}$.

In order to have the least involved sample solution of eq.~(\ref{eq12}), 
we adopted the very conservative choice of $\overline{\eta}_{ext}=1$. In 
Fig.~\ref{fig2} the numerical solution of the above test is shown. In 
plot (a) we could easily see how only after 60 seconds the voltage of 
the capacitor may exceed the value of $0.15\,$volts. Plot (b) tells us 
that the voltage of the capacitor requires some hours to approach 
$0.3\,$volts. Even in the more pessimistic scenario where 
$\overline{\eta}_{in}=10^{-8}$, we see that a macroscopic voltage should 
arise quite rapidly between the plates; see Fig.~\ref{fig3}.

For the sake of completeness, in Fig.~\ref{fig2} and Fig.~\ref{fig3} 
charging profiles for $T=300\,$K and for $T=296\,$K are also shown in 
order to give an hint on how eq.~(\ref{eq12}) behaves with temperature. 
From the experimental point of view this information is important since 
it is hard to maintain the temperature of the environment precisely at 
$T=298\,$K.
 
As can be easily noted, the charging profiles at $T=298\,$K in 
Fig.~\ref{fig2} and Fig.~\ref{fig3} are exactly the same of those in 
Fig.~2 and Fig.~3 of~\cite{io}. All this has a twofold meaning. First of 
all, even with $\eta_{ext}$ five orders of magnitude greater than 
$\eta_{in}$, the counter emission disturbance on the charging process is 
negligible and, as said before, it would become more negligible with the 
suggested choice of $\overline{\eta}_{ext}\ll \overline{\eta}_{in}$, as 
is clear from the mathematics of eq.~(\ref{eq12}).

More important, the physical process firstly introduced in~\cite{io} and 
under study in the present paper is quite robust against the dropping of 
the approximation made in~\cite{io}, namely $\phi_{ext}=\infty$ or 
$\phi_{ext}\gg\phi_{in}$. Moreover, from the experimental point of view, 
the physical process appears to be not so sensible to the specific value 
of $\phi_{ext}$, provided that $\phi_{ext}>1\,$eV.

\section{Discussion}

In this Section we show quantitatively how it is possible to make a 
measurable current flow across the thermo-charged capacitor, once its 
plates are electrically shunted by a suitable resistor, and hence 
allowing the generation of potentially usable voltage, current and 
electric power out of a single thermal source at room temperature (like 
an ordinary battery). In the following, we also try to answer some 
objections, which should naturally and suddenly arise against our 
results.

As explained in Section~2, thermionic emission produces a bias between 
the plates of the spherical capacitor (Figs.~\ref{fig2} and~\ref{fig3}). 
A similar bias also arises across the junction metal/semiconductor 
(since the Ag--O--Cs layer charges up positively) of the inner sphere: 
given the nature of the metal/Ag--O--Cs junction, such a bias has the 
characteristics of a {\em reverse bias}. The reverse bias causes a {\em 
reverse leakage density current} $j_0$, as explained in Section~2, which 
slowly transfers electrons from the inner metallic sphere to the 
Ag--O--Cs layer. This reverse leakage density current $j_0$ is 
microscopic, usually ranging from $10^{-6}$A/cm$^2$ to $10^{-9}$A/cm$^2$ 
for reverse biases of the order of volts or tenths of volts, and its 
intensity is weakly dependent on the magnitude of the reverse bias, 
provided that such a bias is below the breakdown voltage of the 
junction~\cite{oyama,rossi,hsu,dann}. As seen in the previous Section, 
our device is far below that voltage.

Now, the mere fact that a non-zero, almost constant, reverse bias 
leakage density current $j_0$ exists, although very tiny, let the 
thermo-charging process be potentially exploitable. As a matter of fact, 
if we suitably increase the surface area of the inner sphere $S_a$ (and 
also that of the outer one, accordingly), then the contact area between 
the Ag--O--Cs layer and the metal increases. This means that in 
principle it is possible to obtain a macroscopic reverse leakage current 
(RLC) that allows a rather quick transfer of the voltage drop to both 
the terminal leads of the capacitor, since RLC is equal to $j_0S_a$.

\begin{figure}[t]
\centerline{\includegraphics[width=13cm]{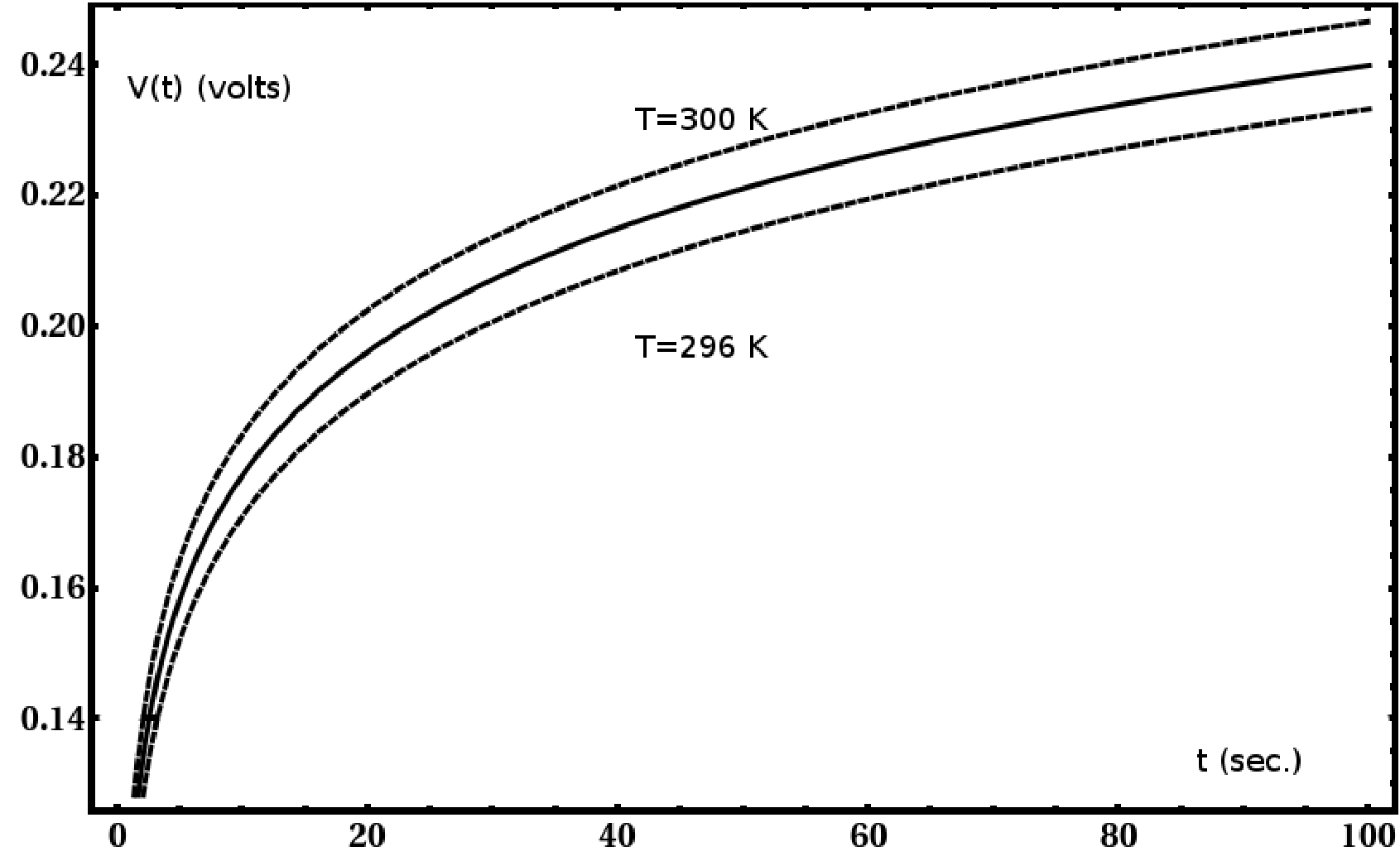}}
\caption{Thermo-charging profile for the spherical capacitor described in 
Section~3 with $\phi_{in}=0.7\,$eV, $\phi_{ext}=4.0\,$eV, $b=2.0\,$m, 
$T=298\,$K, $\overline{\eta}_{in}=10^{-5}$ and $\overline{\eta}_{ext}=1$.
Charging profiles for $T=300\,$K and for $T=296\,$K are also 
shown.}
\label{fig5}
\end{figure}

For the sake of thought experiment, imagine to build a room-sized 
thermo-charged capacitor with radii $a=100\,$cm and $b=200\,$cm. In this 
case the inner sphere surface is equal to $S_a=4\pi a^2\approx 
10^5\,$cm$^2$. Thus, the total reverse leakage current $j_0S_a$ should 
vary between $100\,$mA and $0.1\,$mA; this is a quite macroscopic 
current. The thermo-charging profile for the capacitor with the 
following parameters $\phi_{in}=0.7\,$eV, $\phi_{ext}=4.0\,$eV, 
$b=2.0\,$m, $T=298\,$K, $\overline{\eta}_{in}=10^{-5}$ and 
$\overline{\eta}_{ext}=1$, is represented in Fig.~\ref{fig5}.

Let us now compare the RLC in this case with the thermionic current 
between the spheres $i_{ti}=dQ/dt$ at $t=0\,$s (and $V(0)=0\,$V). It 
should be clear that the value of $i_{ti}$ at $t=0\,$s is the maximum 
value reachable by the thermionic current during the charging process. 
Rearranging eq.~(\ref{eq7}) and eq.~(\ref{eq12}) we obtain,

\begin{equation}
i_{ti}(t)=\frac{dQ}{dt}=\frac{2\pi^2 e b^2}{c^2}
\biggl(\frac{kT}{h}\biggr)^3 \Biggl(\overline{\eta}_{in}\int_{\frac{eV(t) +
\phi_{in}}{kT}}^\infty \frac{x^2 dx}{e^{x}-1} - 4\overline{\eta}_{ext}
\int_{\frac{\phi_{ext}}{kT}}^\infty \frac{x^2 dx}{e^x-1}\Biggr), 
\label{eq13}
\end{equation}
and through numerical calculations for $t=0\,$s we get $i_{ti}(0)\approx 
3.91\times 10^{-10}\,$A.

We note that $i_{ti}\ll j_0S_a$ and this means that the voltage drop 
thermally gained within the plates of the capacitor is quickly 
transferred to both terminal (metallic) leads of the capacitor and can 
be directly detected through an electroscope.

But we can do better. It is possible to directly compare the reverse 
bias leakage density current $j_0$ with the thermionic density current, 
$j_{ti}(0)$, obtained from eq.~(\ref{eq13}) as follows

\begin{equation}
j_{ti}(0)=\frac{i_{ti}(0)}{S_a}=\frac{i_{ti}(0)}{4\pi a^2}
\approx 3.11\times 10^{-15}\,\textrm{A/cm}^2, 
\label{eq14}
\end{equation}
assuming the maximizing condition $a=b/2$ for $dV/dt$ (see eq.~(\ref{eq10})). 

Thus, for $\phi_{in}=0.7\,$eV, $\phi_{ext}=4.0\,$eV, $T=298\,$K, 
$\overline{\eta}_{in}=10^{-5}$, $\overline{\eta}_{ext}=1$ and $j_0$ in 
the reasonable range given before, we see from eq.~(\ref{eq14}) that 
$j_0$ is {\em always} greater that $j_{ti}(0)$: this roughly means that 
the voltage drop thermally gained within the plates of the capacitor is 
{\em always} quickly transferred to both terminal (metallic) leads of 
the capacitor, no matter how big or small is the capacitor. Obviously, 
the bigger is the capacitor, the greater will be the current `flowing' 
through it.

Let us now consider the capacitor of Fig.~\ref{fig1} shunted by a 
suitable resistor $R$. In this case the capacitor behaves like a battery 
which dissipates its power through the resistor (Joule effect).

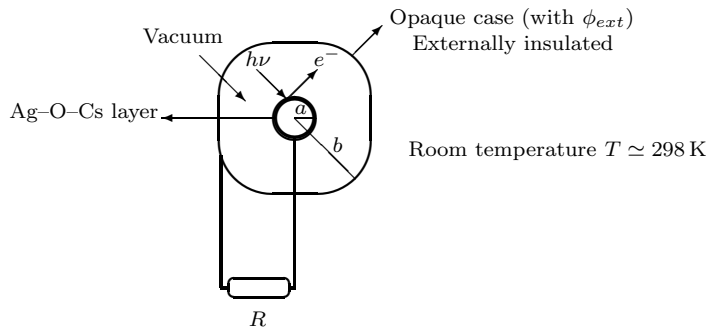
\begin{figure}[t]

\begin{center}
\begin{picture}(150,150)
\setlength{\unitlength}{0.5cm} \thicklines \put(4,6){\circle{1}}
\put(4,6){\circle{1.08}} \put(4,6){\circle{1.1}}
\put(4,6){\oval(4,4)}
\put(4,5.5){\line(0,-1){4}}
\put(4.03,1.5){\line(-1,0){0.2}}
\put(2.06,1.5){\line(1,0){0.2}} 
\put(3.06,1.5){\oval(1.6,0.5)}
\put(2.06,5){\line(0,-1){3.5}}
\put(2.8,0.5){{\scriptsize $R$}} 
\put(6.5,8.5){{\scriptsize Opaque case (with $\phi_{ext}$)}}
\put(7.1,7.8){{\scriptsize Externally insulated}}
\put(0,8){{\scriptsize Vacuum}}
{\thinlines\put(5.5,7.65){\vector(1,1){0.8}}
\put(1.5,7.65){\vector(1,-1){1.2}}} \put(4,6.1){{\scriptsize $a$}}
\put(5,5.1){{\scriptsize $b$}}
{\thinlines\put(4,6){\line(1,0){0.5}}}
{\thinlines\put(4,6){\line(1,-1){1.6}}} \put(7,5){{\scriptsize Room
temperature $T\simeq 298\,$K}} {\thinlines
\put(3.5,6){\vector(-1,0){3}}} \put(-3.5,6){{\scriptsize Ag--O--Cs
layer}} {\thinlines \put(3,7.3){\vector(1,-1){0.8}}} {\thinlines
\put(3.8,6.5){\vector(1,1){0.8}}} \put(2.7,7.4){\scriptsize $h\nu$}
\put(4.5,7.4){\scriptsize $e^-$}


\end{picture}
\end{center}
\caption{Thermo-charged spherical capacitor shunted by a resistor $R$.}
\label{fig6}
\end{figure}

Consider the electrical circuit depicted in Fig.~\ref{fig6}. In steady 
state conditions, voltage drop $V_s$ and current $i_s$ should be the 
same across the capacitor and the resistor; moreover, they should be 
constant in time. According to Ohm's Law, we must have $R=V_s/i_s$, and 
this relation also gives the numerical value of the resistance $R$ 
needed to have these particular values of $V_s$ and $i_s$.

Given the equations of the thermo-charged capacitor described before, we 
must have,

\begin{equation}
i_{s}=\frac{2\pi^2 e b^2}{c^2}
\biggl(\frac{kT}{h}\biggr)^3\Biggl(\overline{\eta}_{in}\int_{\frac{eV_s +
\phi_{in}}{kT}}^\infty \frac{x^2 dx}{e^{x}-1} - 4\overline{\eta}_{ext}
\int_{\frac{\phi_{ext}}{kT}}^\infty \frac{x^2 dx}{e^x-1}
\Biggr). 
\label{eq15}
\end{equation}

The power $P_s$ provided by the thermo-charged capacitor is calculated 
as,

\begin{equation}
P_s=V_si_{s}=\frac{2\pi^2 e b^2V_s}{c^2}
\biggl(\frac{kT}{h}\biggr)^3\Biggl(\overline{\eta}_{in}\int_{\frac{eV_s +
\phi_{in}}{kT}}^\infty \frac{x^2 dx}{e^{x}-1} - 4\overline{\eta}_{ext}
\int_{\frac{\phi_{ext}}{kT}}^\infty \frac{x^2 dx}{e^x-1}
\Biggr). 
\label{eq16}
\end{equation}
 
The power per unit surface of the inner sphere ${\cal P}_s$ is then, 

\begin{equation}
{\cal P}_s=\frac{V_si_{s}}{S_a}=\frac{2\pi eV_s}{c^2}
\biggl(\frac{kT}{h}\biggr)^3\Biggl(\overline{\eta}_{in}\int_{\frac{eV_s +
\phi_{in}}{kT}}^\infty \frac{x^2 dx}{e^{x}-1} - 4\overline{\eta}_{ext}
\int_{\frac{\phi_{ext}}{kT}}^\infty \frac{x^2 dx}{e^x-1}
\Biggr). 
\label{eq16}
\end{equation}

\begin{figure}[t]
\centerline{\includegraphics[width=13cm]{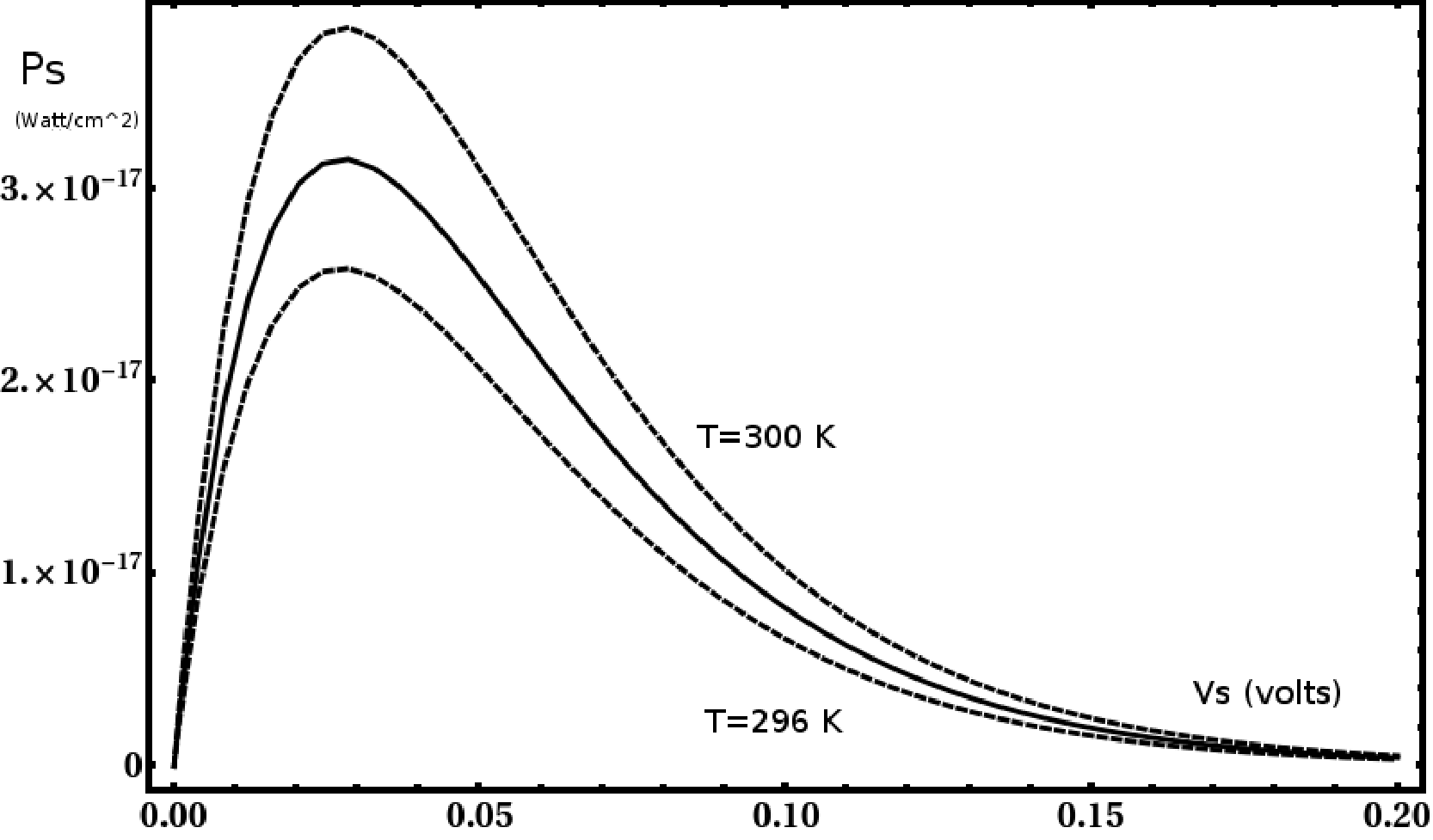}}
\caption{Power output per unit surface area of the inner sphere, ${\cal P}_s$,
against voltage drop $V_s$ of the electrical circuit 
capacitor/resistor depicted in Fig.~\ref{fig6}. Power outputs for $T=300\,$K 
and for $T=296\,$K are also shown.}
\label{fig7}
\end{figure}

In Fig.~\ref{fig7} the power ${\cal P}_s$ is plotted against the steady 
state voltage $V_s$. For the capacitors described in this paper, namely 
that with $a=10\,$cm, $S_a\approx 1260\,$cm$^2$ and that with 
$a=100\,$cm, $S_a\approx 10^5\,$cm$^2$, we obtain $P_{max}\approx 
4\times 10^{-14}\,$Watts and $P_{max}\approx 4\times 10^{-12} \,$Watts, 
respectively. These are quite microscopic power outputs indeed, 
considering further that the second capacitor described has 
`uncomfortable' room-sized dimensions.

There are now few doubts that one of the definite results of our paper 
is that the thermo-charged capacitors described here are highly 
`ineffective'. Anyway, it is important to experimentally test their 
functioning, since if they work according to the analysis done in this 
paper, then we would have a {\em reproducible} Second Law violation: 
therefore, we believe that its {\em smallness} is a secondary problem 
(that can be overcome with further future research), provided that this 
{\em smallness} is not to forbid a clear and unambiguous result with 
confounding environmental factors.

We have seen that the current and the power outputs of circuits like 
those depicted in Fig.~\ref{fig6} are microscopic, in particular if the 
physical dimensions of the spheres are centimetric. Nevertheless, the 
voltage drop of non-shunted capacitors is macroscopic (of the order of 
$0.1\,$V), already for a single capacitor: thus, it should not be 
difficult to build tens of centimetre-sized vacuum capacitors, wired in 
series, so as to produce a voltage drop of volts or tens of volts. This 
output is far from the risk to be ambiguously interpreted.

From an experimental perspective, some manufacturing difficulties have 
to be faced. For example, stable ultra-high vacuum required inside the 
capacitor may pose a technical challenge, mainly for metre-sized device.  
Nonetheless, vacuum technology is currently quite sophisticated and 
mature and we expect that the construction of light, vacuum-proof 
centimetre-sized capacitors should not pose any concern at all.

\begin{figure}[t]
\centerline{\includegraphics[height=5.3cm,width=7cm]{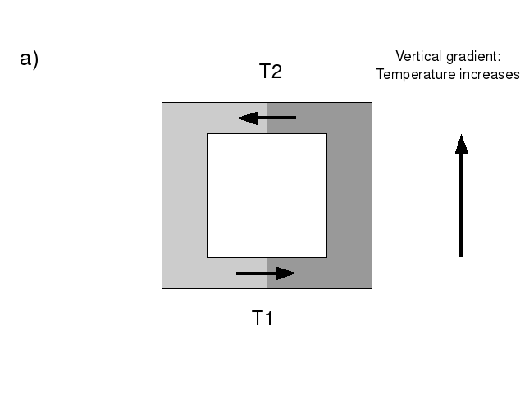}
  \includegraphics[height=5.3cm,width=7cm]{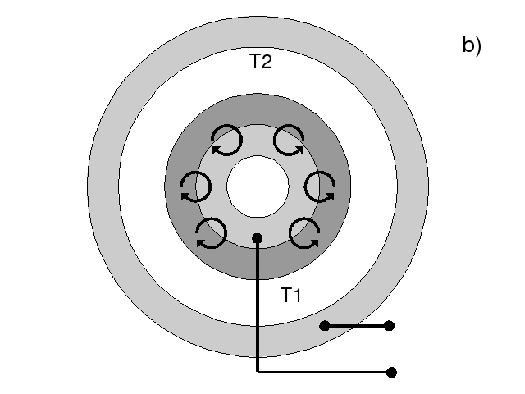}}
\caption{$T_2>T_1$. a) sketch of a classical shunted thermocouple, b) 
a sketched section of our capacitor. The arrows show the direction of 
currents.}
\label{fig8}
\end{figure}

One issue that may be experimentally important is how the naturally 
present spatial thermal gradient\footnote{Like that usually present 
inside a room between floor and ceiling.} affects the functioning of our 
capacitor. The inner sphere being made of two materials with different 
work functions in contact, one may expect the presence of the well-known 
{\em thermocouple} (voltage/current) effect (see Fig.~\ref{fig8}-a), 
that could make difficult and ambiguous the verification of the 
phenomenon described in this paper.

As a matter of fact, the geometry of our capacitor is such that a 
macroscopic thermal gradient across the inner sphere should not 
significantly affect the verification of our results: as shown in 
Fig.~\ref{fig8}-b any possible thermocouple current remains confined 
inside the inner sphere, and the thermocouple voltage is zero being the 
thermocouple `shunted'. This confounding factor could be greatly reduced 
if it were possible to perform the experiment in micro-gravity 
environment, like, for example, that on International Space Station 
(ISS). Moreover, the use of many small (centimetre-sized) capacitors, 
rather than a single metre-sized one, greatly reduces the problem.

For what concerns another possible disturbing effect, the Thomson 
effect\footnote{Voltage/current generation due to temperature gradient 
within a system made of a single material.}, one can imagine to build an 
identical copy of our capacitor without the Ag--O--Cs coating on the 
inner sphere and to place it in the same environment of our main 
capacitor, side by side. In such a way, one can use the (possible) 
output of the ``uncoated'' capacitor, due to the Thomson effect, to 
cancel out the disturbance in our main capacitor due to the same effect.

Finally, let us now anticipate (and try to exhaustively respond to) some 
possible critics. Among the main objections to our results, probably the 
first one is that our device appears to suffer the same shortcomings of 
the self-rectifying diode scheme (see, for example, 
\cite[Brillouin]{brill} and \cite[McFee]{mcf} and the references 
therein). We have to stress that our device is quite different from a 
solid-state diode: in solid-state diodes the physical contact, through 
the n--p junction, between the two terminals and the dynamic balance 
between the built-in electric field and diffusion forces across the 
junction prevent the establishment of a non-random, steady charge 
movement far from the depletion region toward the terminals, and thus 
the creation of a voltage drop between the terminal leads.

In our device, the presence of vacuum between the outer sphere and the 
inner one (there is no physical contact inside the capacitor), and the 
fact that only one sphere is covered with a low work function material, 
do not allow the above dynamic balance, and a definite, one-way 
migration of charges across the plates (to the terminal leads) should be 
possible, as described above.

\section{Electro-mechanical analogue of a thermo-charged capacitor}

Thermo-charged capacitor has an easily understandable electro-mechanical 
analogue, which is both instructive and explicative and thus worthwhile 
to describe.

\begin{figure}[t]
\centerline{\includegraphics[height=6.5cm,width=9cm]{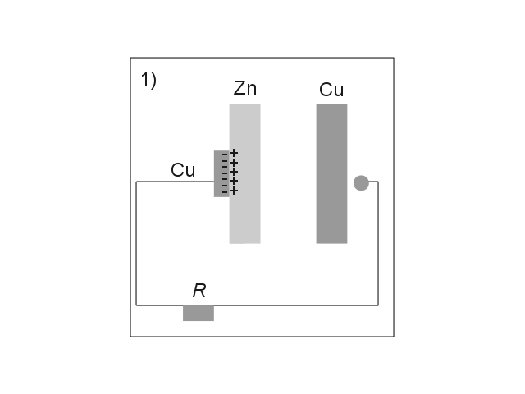}
  \includegraphics[height=6.5cm,width=9cm]{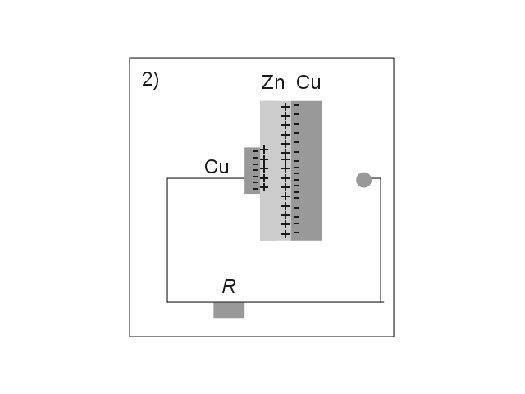}}
\centerline{\includegraphics[height=6.5cm,width=9cm]{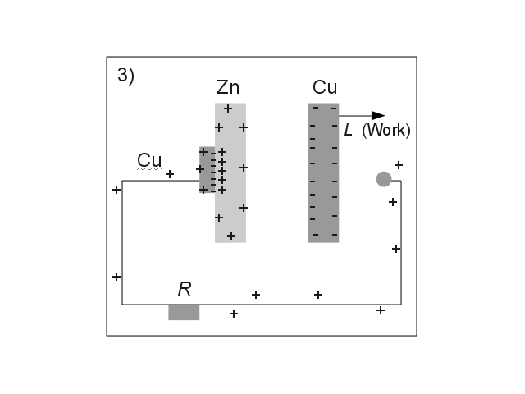}
  \includegraphics[height=6.5cm,width=9cm]{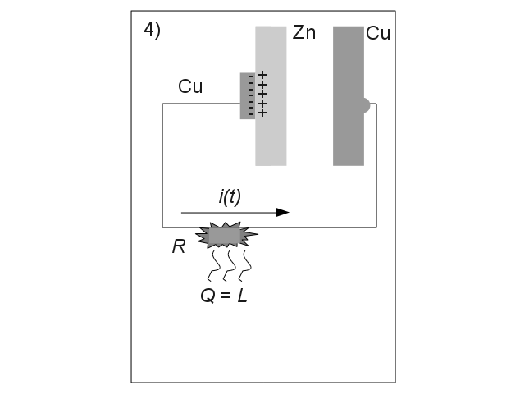}}
\caption{Electro-mechanical analogue of a thermo-charged capacitor.}
\label{fig9}
\end{figure}

Consider the device depicted in Fig.~\ref{fig9}-1. It is essentially a 
parallel plates capacitor with one plate made of a metal with relatively 
low work function, e.g. zinc (Zn), and the other one made of a 
relatively high work function metal, e.g. copper (Cu). This last plate 
is also free to move in the space. Moreover, a copper wire (with a load 
$R$) is connected to the zinc plate through a small junction Cu-Zn. The 
Cu-Zn junction generates a very thin (the junction being a metal to 
metal one) depletion layer along the contact surface, where positive and 
negative charges are localized after the dynamically balanced drift of 
electrons from zinc to copper through the contact area 
(Fig.~\ref{fig9}-1).

The first step of the electro-mechanical cycle we are going to describe 
consists in moving the copper plate toward the zinc one until the 
contact. In this phase no significant external work is required. After 
the contact, a second (and greater) depletion layer forms between zinc 
and copper plates. Also in this case, the new Cu-Zn junction generates a 
very thin depletion layer along the new (and wider) contact surface 
(Fig.~\ref{fig9}-2).

In step two, an external work $L$ is applied to the copper plate in 
order to remove it from the zinc plate: $L$ is different from zero since 
the previous charges displacement across the new Cu-Zn junction makes 
the two plates attract each other. When the two plates are again 
suitably removed, the charges, initially localized within the thin 
depletion layer, are free to spread across the surfaces of the two 
metallic plates and wire, satisfying equi-potentiality 
(Fig.~\ref{fig9}-3; see, for example, \cite{cott}).

In the third step of our cycle, we put the negatively charged copper 
plate and the positively charged copper wire into contact 
(Fig.~\ref{fig9}-4). As soon as the contact is made, electrons start to 
flow from copper plate to zinc plate across the copper wire/load until 
both plates become neutral. Due to the Joule effect in the discharge, 
the load $R$ heats up. The amount of heat $Q$ transferred to the 
environment is nearly equal to the external work $L$ done to the system 
in step number two (First Law of Thermodynamics).

The fourth step, that closes the cycle, is trivial and consists in 
moving the copper plate to its initial position, as in 
Fig.~\ref{fig9}-1.

The analogy between this scheme and the thermo-charged capacitor should 
be clear: both devices work with materials having different work 
functions, and in both devices current (electrons) flows across contact 
junctions to re-establish electrical equilibrium.

The main difference is in the source of the energy needed to make the 
current flow. In the electro-mechanical scheme the source is the 
external work $L$ done to the system through the movable copper plate. 
In the thermo-charged capacitor it is the black-body radiation (of the 
uniformly heated environment) which provides the kinetic energy to 
electrons and let them fly to the external (fixed) plate of the 
capacitor.

\section*{Acknowledgements}

This work has been partially supported by the Italian Space Agency under 
ASI Contract No.~1/015/07/0. The author thanks the three anonymous 
reviewers for their comments and suggestions.

\end{document}